\documentclass[preprint]{aastex}
\usepackage{aas_symbols}
\usepackage{natbib}

\slugcomment{Accepted for publication in the {\sl Astronomical Journal}}
\shorttitle{Cepheid diameters}
\shortauthors{Armstrong et al.}

\begin{document}

\title{Diameters of $\delta$ Cephei and $\eta$ Aquilae Measured with the Navy Prototype Optical 
Interferometer}

\author{J.~T.\ Armstrong\altaffilmark{1}, Tyler~E.~Nordgren\altaffilmark{2}, M.~E.~Germain\altaffilmark{2}, 
Arsen~R.~Hajian\altaffilmark{3}, R.~B.~Hindsley\altaffilmark{1,3}, C.~A.~Hummel\altaffilmark{3}, 
D.~Mozurkewich\altaffilmark{1}, \and R.~N.~Thessin\altaffilmark{4}}

\altaffiltext{1}{Remote Sensing Division, Naval Research Laboratory, 4555 Overlook Ave.\ SW, Washington, DC 
20375; tarmstr, mozurk@rsd.nrl.navy.mil}
\altaffiltext{2}{Astrometry Department, US Naval Observatory; postal address: US Naval Observatory, P.O.\ Box 
1149, Flagstaff, AZ 86002; nordgren@nofs.navy.mil, meg@sextans.lowell.edu}
\altaffiltext{3}{Astrometry Department, US Naval Observatory, 3450 Massachusetts Ave.\ NW, Washington, DC 
20392-5420; hajian, rbh, cah@gemini.usno.navy.mil}
\altaffiltext{4}{MSC 890, California Institute of Technology, Pasadena, CA  91126; rthessin@its.caltech.edu}

\begin{abstract}
We have measured the diameters of the Cepheid variables $\delta$~Cephei (18 nights) and $\eta$~Aquilae (11 
nights) with the Navy Prototype Optical Interferometer.  The primary results of these observations are the 
mean angular diameters $\langle\theta_{\rm LD}\rangle$ of these Cepheids: $1.520 \pm 0.014$ milliseconds of 
arc (mas) for $\delta$~Cep and $1.69 \pm 0.04$~mas for $\eta$~Aql.  We also report limb-darkened diameters for 
the check stars in this program: for $\beta$~Lac, $\theta_{\rm LD} = 1.909 \pm 0.011$~mas, and for 12~Aql, 
$\theta_{\rm LD} = 2.418 \pm 0.010$~mas.  When combined with radius estimates from period-radius relations in 
the literature, the Cepheid angular diameters suggest distances slightly smaller than, but still consistent 
with, the Hipparcos distances.  Pulsations are weakly detected at a level of $\sim 1.5\sigma$ to $2\sigma$ for 
both Cepheids.

\end{abstract}


\section{Introduction}\label{sect:intro}
Cepheid variables represent the first step on the extragalactic distance scale.  But precise measurements of 
the distances to individual Cepheids are difficult.  Trigonometric parallax uncertainties for even the nearest 
Cepheids are large: the Hipparcos parallaxes $\pi_{\rm Hip}$ are $3.32 \pm 0.58$~mas for $\delta$~Cep and 
$2.78 \pm 0.91$~mas for $\eta$~Aql \citep{ESA97}.  The Baade-Wesselink and similar methods use the surface 
brightness $S$ and apparent magnitude as a function of pulsation phase $\phi$ to estimate the variation in the 
ratio $R(\phi)/\langle R \rangle$, where $R$ is the radius, and use the integrated radial velocity curve to 
estimate $R(\phi) - \langle R \rangle$.  The combination gives both $R(\phi)$ and $\langle R \rangle$, while 
the distance is derived from $\langle R \rangle$ and $S$.

The optical interferometers now coming into operation offer the possibility of measuring the angular diameter 
variation directly, and of combining that measurement with the integrated radial velocity result to derive the 
distance.  Even if the precision of the diameter measurements is insufficient to show the variation, as in 
these data, the diameters can be used to test whether the Cepheids conform to the surface brightness vs.\ 
color relations for nonpulsating stars.

The first interferometric determination of the diameter of a Cepheid is that of \citet{Mou97}.  They reported 
a mean limb-darkened diameter of $1.63 \pm 0.19$~mas for $\delta$~Cep, but the precision of the diameters on 
individual nights was insufficient to detect the pulsation.  We report here on measurements of $\delta$~Cep 
($P = 5.366316$~d; \citealt{Mof85}) and $\eta$~Aql ($P = 7.176726$~d; \citealt{Sza91}) with the Navy Prototype 
Optical Interferometer (NPOI).\footnote{The Navy Prototype Optical Interferometer is a joint project of the 
Naval Research Laboratory and the U.S.\ Naval Observatory, in cooperation with Lowell Observatory, and is 
funded by the Office of Naval Research and the Oceanographer of the Navy.} Preliminary results for these stars 
and for the check star $\beta$~Lac were reported by \citet{Nor99}.  The precision of these measurements is an 
improvement over the previous results; however, the pulsation is only weakly seen, and the distances derived 
from the amplitude of the diameter variation are of low precision.  Even so, the mean diameters, combined with 
the radii derived from the surface brightness and radial velocity curves, can produce distance estimates.  
Mean diameters of the Cepheids $\zeta$~Gem and $\alpha$~UMi, with comparisons of the four NPOI-measured 
Cepheids to period-radius and period-mass relations in the literature, are given in a companion paper 
\citep{Nor00}.

\section{Observations}\label{sect:obs}
We observed $\delta$~Cep and $\eta$~Aql with the east, center, and west (E, C, W) elements of the astrometric 
array of the NPOI between 1997 July and 1998 October.  The baseline lengths were 37.5~m, 22.2~m, and 18.9~m at 
azimuths of $-67\fdg 5$, $63\fdg 6$, and $86\fdg 0$ (EW, CW, and CE).  Each baseline produces squared fringe 
visibilities $V^{2}_{\lambda}$ in 32 channels covering the $\lambda\lambda 850-450$~nm range, although the 
sensitivity blueward of $\sim\lambda 550$~nm is low.  For the results reported here, we used data in the 
reddest 13 channels, covering the $\lambda\lambda 850-600$~nm range.  We removed a small number of spectral 
channels with bad detectors, as well as the channel contaminated by the HeNe delay-line metrology laser.  The 
NPOI is described in detail by \citet{Arm98}.

We chose an unresolved star close in position to each Cepheid to act as a visibility calibrator: $\alpha$~Lac 
with an estimated diameter of 0.5~mas for $\delta$~Cep, and $\lambda$~Aql (0.5~mas) for $\eta$~Aql.  In 
addition, we observed $\beta$~Lac, a $\sim 1.9$~mas star $6\fdg 3$ from $\delta$~Cep, and 12~Aql, a $\sim 
2.4$~mas star $9\fdg 9$ from $\eta$~Aql, as check stars for observations in 1998 July and later.  Each night's 
observing list consisted of $\sim 10-15$ stars, including the Cepheids, their calibrators, and the check 
stars.  We placed each Cepheid, its calibrator, and its check star consecutively in the observing list to 
minimize changes in observing conditions during the sequence.  Typical intervals between successive 90~s scans 
on the same star were $15-45$~min.  The check star--calibrator separations, $\beta$~Lac--$\alpha$~Lac and 
12~Aql--$\lambda$~Aql are $2\fdg 2$ and $1\fdg 1$, respectively; however, the Cepheid--calibrator separations, 
$\delta$~Cep--$\alpha$~Lac and $\eta$~Aql--$\lambda$~Aql, are considerably larger, at $8\fdg 1$ and $8\fdg 9$, 
respectively.  In 
Tables~1 and 2, we list the date of observation, the mean 
phase, the number of scans, the diameter, and the estimated error for each night (see \S\ref{subsect:uncert} 
for a discussion of the determination of the uncertainties).  For the two Cepheids, 
Tables~1
and 2 also show the best-fit mean diameters to a pulsating model, assuming a phase shift 
$\Delta\phi_v = 0$ between variations in the NPOI diameters and variations derived from the radial velocities 
(see \S\ref{sect:pulse}).  These diameters are not the means of the measured diameters because our phase 
coverage is not uniform.


\begin{table}
    \label{table:delcep}
\begin{tabular}{lcccl}
\multicolumn{5}{c}{T{\sc able} 1. L{\sc imb-darkened}}    \\
\multicolumn{5}{c}{\sc angular diameters:}                \\
\multicolumn{5}{c}{$\delta$ C{\sc ep and} $\beta$ L{\sc ac}} \\[10pt]
\tableline\tableline                                      \\[-8pt]
JD\hspace{0.2in}    &       & No. & $\theta_{\rm LD}$ & $\sigma_{\theta}$ \\
$-2450000$          & $\langle\phi\rangle$\tablenotemark{a} & Scans & (mas) & (mas) \\
\tableline                                                \\[-2pt]
\multicolumn{5}{c}{$\delta$ Cep}                          \\[6pt]
\tableline                                                \\[-2pt]
\phantom{1}788.615 & 0.764  &  \phn\phn 8 & 1.51  & 0.05  \\
\phantom{1}994.903 & 0.206  &  \phn\phn 3 & 1.74  & 0.10  \\
\phantom{1}995.934 & 0.398  &  \phn    11 & 1.48  & 0.04  \\
\phantom{1}996.967 & 0.590  &  \phn\phn 4 & 1.55  & 0.05  \\
\phantom{1}997.929 & 0.770  &  \phn\phn 8 & 1.39  & 0.08  \\
\phantom{1}998.931 & 0.956  &  \phn\phn 4 & 1.36  & 0.17  \\
          1007.965 & 0.640  &  \phn\phn 4 & 1.52  & 0.10  \\
          1008.931 & 0.820  &  \phn\phn 4 & 1.45  & 0.09  \\
          1009.916 & 0.003  &  \phn\phn 9 & 1.51  & 0.03  \\
          1010.917 & 0.190  &  \phn\phn 9 & 1.61  & 0.11  \\
          1011.910 & 0.375  &  \phn\phn 6 & 1.61  & 0.15  \\
          1012.904 & 0.560  &  \phn\phn 9 & 1.56  & 0.12  \\
          1088.807 & 0.705  &  \phn\phn 2 & 1.38  & 0.09  \\
          1089.767 & 0.883  &  \phn\phn 6 & 1.52  & 0.09  \\
          1093.738 & 0.623  &  \phn    19 & 1.51  & 0.05  \\
          1096.765 & 0.187  &  \phn\phn 1 & 1.61  & 0.19  \\
          1097.778 & 0.376  &  \phn\phn 9 & 1.58  & 0.04  \\
          1098.828 & 0.572  &  \phn\phn 4 & 1.59  & 0.09  \\
                   &        &             &       &       \\
\multicolumn{2}{l}{Total scans} &  120    &       &       \\
\multicolumn{3}{l}{Best-fit diameter\tablenotemark{b}} & 1.520 & 0.014 \\
\tableline\\[-2pt]
\multicolumn{5}{c}{$\beta$ Lac}                           \\[6pt]
\tableline                                                \\[-2pt]
all                &  ---   &         114 & 1.909 & 0.011 \\
\tableline\\

\end{tabular}
\tablenotetext{a}{From epoch = JD~2443674.144, $P = 5.366316$~days\\ 
(Moffett \& Barnes 1985)}
\tablenotetext{b}{From model fit with phase shift $\Delta\phi_v = 0$; see \S\ref{sect:pulse} \\
and Table~\ref{table:models}}

\end{table}


     
\begin{table}
    \label{table:etaaql}
\begin{tabular}{lcccl}
\multicolumn{5}{c}{T{\sc able} 2. L{\sc imb-darkened}}\\
\multicolumn{5}{c}{\sc angular diameters}             \\
\multicolumn{5}{c}{$\eta$ A{\sc ql and} 12 A{\sc ql}} \\[10pt]
\tableline\tableline                                  \\[-8pt]
JD\hspace{0.2in} &         & No. & $\theta_{\rm LD}$ & $\sigma_{\theta}$ \\
$-2450000$         &$\langle\phi\rangle$\tablenotemark{a}     & Scans  & (mas) & (mas) \\
\tableline                                            \\[-2pt]     
\multicolumn{5}{c}{$\eta$ Aql}                        \\[6pt]
\tableline                                            \\[-2pt]
\phantom{1}638.863 & 0.997  & \phn 4 & 1.92  & 0.13   \\
\phantom{1}640.895 & 0.280  & \phn 3 & 1.73  & 0.11   \\
\phantom{1}641.858 & 0.414  & \phn 4 & 1.91  & 0.13   \\
\phantom{1}997.824 & 0.014  & \phn 3 & 1.65  & 0.12   \\
\phantom{1}998.885 & 0.162  & \phn 6 & 1.63  & 0.10   \\
          1007.883 & 0.416  & \phn 3 & 1.92  & 0.10   \\
          1008.915 & 0.559  & \phn 2 & 1.45  & 0.28   \\
          1009.855 & 0.690  & \phn 7 & 1.62  & 0.07   \\
          1010.850 & 0.829  & \phn 6 & 1.48  & 0.07   \\
          1011.825 & 0.965  & \phn 6 & 1.63  & 0.06   \\
          1012.861 & 0.109  & \phn 4 & 1.63  & 0.10   \\
                   &        &        &       &        \\
\multicolumn{2}{l}{Total scans} & 48 &       &        \\
\multicolumn{3}{l}{Best-fit diameter\tablenotemark{b}}   & 1.69  & 0.04  \\
\tableline                                            \\[-2pt]
\multicolumn{5}{c}{12 Aql}                            \\[6pt]
\tableline                                            \\[-2pt]
all                &  ---   &     29 & 2.418 & 0.010  \\
\tableline                                            \\
\end{tabular}
\tablenotetext{a}{From epoch = JD~2442794.726, $P = 7.176726$~days\\ 
(Szabados 1991)}
\tablenotetext{b}{From model fit with phase shift $\Delta\phi_v = 0$; see \S\ref{sect:pulse}\\
and Table~\ref{table:models}}
\end{table}


\section{Data Reduction}\label{sect:reduction}

\subsection{Angular Diameters}\label{subsect:angdiam}

We modeled the Cepheids and the check stars as limb-darkened disks, using \citep{Bro74}
\begin{equation}
    V^{2}\left({\theta_{\rm LD}, |{\bf u}_\lambda|}\right) = 
    \left({\alpha \over 2} + {\beta \over 3}\right)^{-2}
    \left[
       {{\alpha J_{1}(\pi \theta_{\rm LD} |{\bf u}_\lambda|)}\over{\pi \theta_{\rm LD} |{\bf u}_\lambda|} }
        + {{\beta (\pi/2)^{1/2}J_{3/2}(\pi \theta_{\rm LD} |{\bf u}_\lambda|)}\over
	{(\pi \theta_{\rm LD} |{\bf u}_\lambda|)^{3/2}}}
    \right]^{2},
    \label{eq:v2LD}
\end{equation}
where ${\bf u}_\lambda$ is the projected interferometer baseline vector in the $u,v$ plane at wavelength 
$\lambda$, and where $\alpha = 1- x_{\lambda}$ and $\beta = x_{\lambda}$, with $x_{\lambda}$ as the linear 
limb-darkening coefficient.  $J_1$ and $J_{3/2}$ are the Bessel functions of the first kind and of the first 
and three-halves orders.  We then fit $\theta_{\rm LD}$ to the data from each night using Eq.~(\ref{eq:v2LD}) 
and obtained a preliminary uncertainty estimate from the fit.

We took the limb darkening coefficients from \citet{VHa93}, where they are tabulated as functions of $T_{\rm 
eff}$ and $\log g$.  For the Cepheids, we estimated $\log g$ as a function of phase from the mean $\log g$ of 
\citet{Hin89}, modified by the acceleration inferred by the radial velocity curves of \citet{But93} and 
\citet{Sha58} for $\delta$~Cep and the radial velocity curves of \citet{Eva76} and \citet{Jac81} for 
$\eta$~Aql.  For the check stars, we took $\log g = 2.97$ for $\beta$~Lac and $\log g = 2.75$ for 
12~Aql \citep{McW90}.

The route to an estimate of $T_{\rm eff}$ as a function of phase for the Cepheids is a bit more roundabout.  
We used the $T_{\rm eff}$ vs.\ $(V-K)_{0}$ relation of \citet{Bla94}.  The $(V-K)_{0}$ data were derived from 
spline fits to the $V$ and $K$ photometry of \citet{Bar97}, from which we generated $(V-K)$ at the phases of 
our observations.  To deredden $V$, we used $A_{\rm V}/E(B-V) = R_{\rm V} = 3.46$ \citep{Eva93} with $E_{B-V} 
= 0.092$ for $\delta$~Cep and 0.149 for $\eta$~Aql.  To deredden $K$, we used $A_{\rm K}/E(V-K) = R_{\rm K} = 
0.37$ \citep{Eva93} with $E_{V-K} = 0.28$ for $\delta$~Cep and 0.46 for $\eta$~Aql.  For the check stars, we 
took $T_{\rm eff} = 4710$ for $\beta$~Lac and $T_{\rm eff} = 4600$ for 12~Aql \citep{McW90}.

Neither the choice of $T_{\rm eff}$ nor of $\log g$ is critical.  The derived diameter is only weakly dependent 
on temperature ($d\theta_{\rm LD}/dT \approx 0.01$~$\mu$as/deg between 5000~K and 7000~K) or gravity 
($d\theta_{\rm LD}/d(\log g) \approx 4.5$~$\mu$as per unit change in $\log g$).

We estimated the mean diameter for each Cepheid by fitting the set of nightly diameters to a pulsation model 
whose shape was derived from radial velocity data, as described in \S\ref{sect:pulse}.  The $\langle 
\theta_{\rm LD}\rangle$ results reported here are those for models in which the phase shift $\Delta\phi_v$ 
between the diameter data and the radial velocity data is fixed at zero.

\subsection{Calibration}\label{subsect:calib}

Because the Cepheids are only marginally resolved with the array configuration used here, and since the 
pulsation amplitudes are expected to be only $\sim 10$\%, the accuracy of the calibration and reduction are 
critical.  We took two approaches to calibrating $V^{2}_{\lambda}$ to investigate the dependence of the 
results on the calibration technique.

One approach was to treat all nights the same.  Under this approach, we calibrated our data using three 
related techniques.  The most direct was to use the calibrator scan nearest in time to calibrate each Cepheid 
scan.  In one refinement, we used a linear interpolation in time between the calibration scans just before and 
after the Cepheid scan.  In a second, we used Gaussian smoothing of the calibration correction with a width of 
40 minutes.

The second approach was to calibrate each night differently, keeping the following techniques at our disposal: 
Gaussian smoothing of the calibration correction, with the smoothing width adjustable over a range of 10 to 80 
minutes; fitting a quadratic function of time to the calibrator scans; and fitting a linear function of fringe 
delay jitter to the calibrator scans.  We tried this approach twice: once using only one of these three 
techniques for any given night, and once using the combination of them that seemed appropriate night by night.

The standard deviation of the mean diameters between calibrations was about $0.02$~mas.  For the results 
reported here, we averaged the diameters determined with the various calibrations.

We also varied the bluest spectral channel used in the reduction: increasing the minimum wavelength avoids 
using the noiser data from the bluest channels, but at the cost of reducing the maximum spatial frequency 
sampled.  We found that the choice of minimum wavelength had very little effect.  Tests with one night's data 
indicated a variation in the derived diameter of $\delta$~Cep of only $0.0086~$mas between using 13 channels 
($850-600$~nm) and using only four channels ($850-770$~nm); between using 13 channels and 10 channels 
($850-650$~nm), the difference was $0.0005~$mas.

\subsection{Diameter Uncertainties}\label{subsect:uncert}

Estimating the uncertainties in the nightly diameter measurements of the Cepheids and of the mean diameters of 
both the Cepheids and the check stars is complicated by the fact that they are dominated not by the 
signal-to-noise ratio within a given scan, but by variations due to the fact that the calibrator data are 
taken at different times at different positions on the sky than the Cepheid data.  The size of these 
variations must be measured from the data themselves.

The check star data should be the source of that measurement.  But we found that the variations in the Cepheid 
diameters within a night are larger than those in the check stars.  (The Cepheid diameter changes within a 
night due to pulsation are smaller than the uncertainties.)  There are two reasons for this discrepancy.  The 
first is that the Cepheids are smaller than the check stars ($\sim 1.5$ and 1.7~mas for $\delta$~Cep and 
$\eta$~Aql, versus $\sim 1.9$ and 2.4~mas for $\beta$~Lac and 12~Aql).  The derivative $dV^2/d\theta$ is 
smaller for small stars, so a given error in $V^2$ leads to a larger error in $\theta$.  The size of this 
effect can be calculated: the errors in $\theta$ are 30\% larger for $\delta$~Cep than for $\beta$~Lac, and 
55\% larger for $\eta$~Aql than for 12~Aql.

The second reason is that, as noted above, the check stars are closer to the calibrators than are the 
Cepheids, and should therefore be better calibrated.  This effect is harder to quantify, but the residuals 
after subtracting the best-fit pulsation models (see \S\ref{sect:pulse}) from the calibrated data suggest 
an additional increase in diameter uncertainties of $\sim 1.7$ and $\sim 5$ times for $\delta$~Cep 
and $\eta$~Aql, respectively, than in their check stars.

Instead of trying to estimate the Cepheid diameter uncertainties from the check star uncertainties, we chose 
the following method.  We started with the estimated uncertainty of the diameter fit to each night's data.  We 
added the standard deviation of the mean diameters among calibration techniques in quadrature.  Finally, we 
scaled the uncertainties thus derived to produce a reduced $\chi^2$ of unity for a fit to the best-fit 
pulsation models.  These uncertainties, which appear in 
Tables~1 and 2, thus 
represent an internally consistent set of error estimates, while the night-to-night variations give a relative 
measure of the quality of the data.  We chose a similar strategy for the check star diameters, scaling their 
uncertainties to produce a reduced $\chi^2$ of unity for a fit to a constant diameter.

\begin{figure}
    \epsscale{.8}
    \figurenum{1}
    \plotone{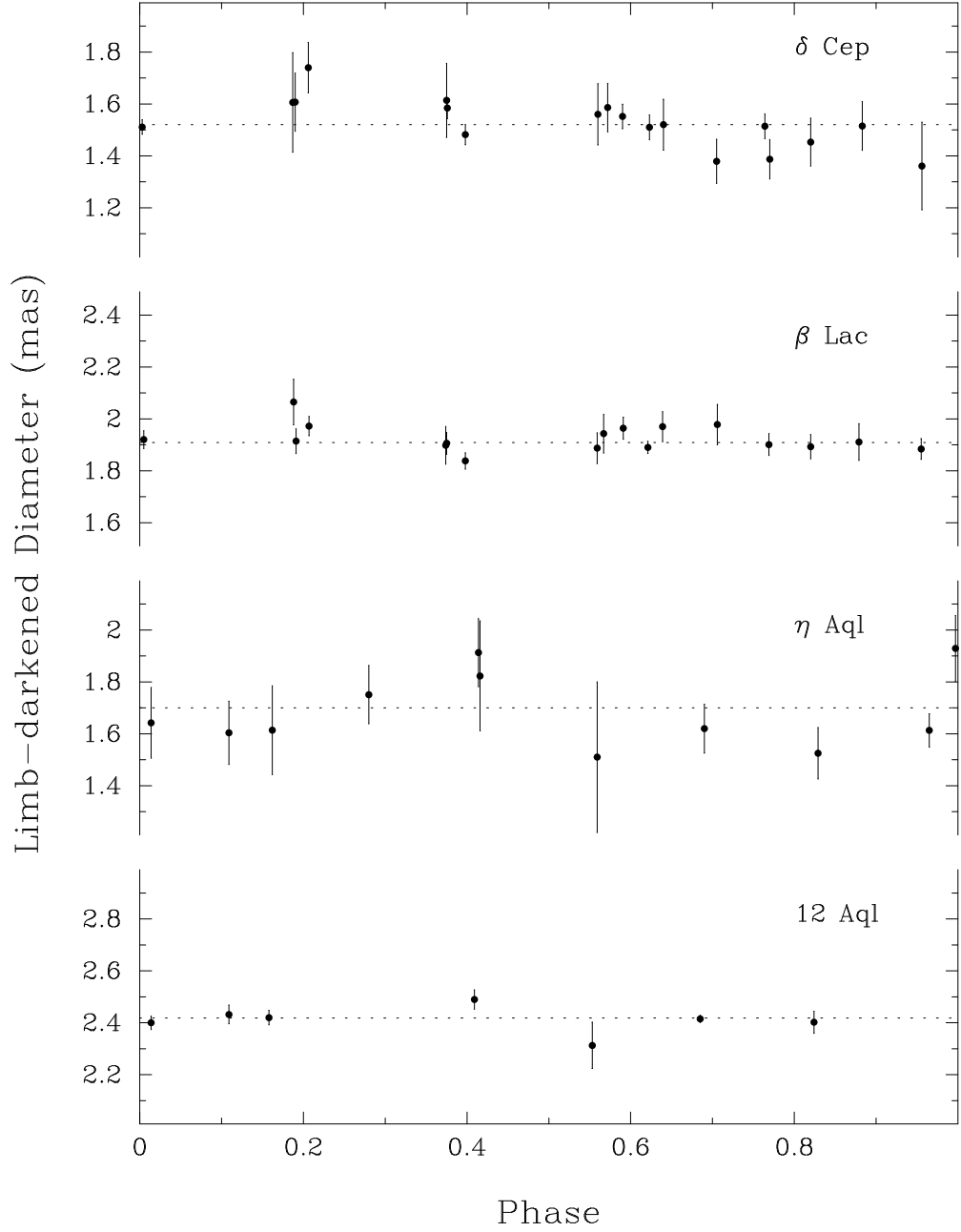}
    \caption
    {Limb darkened angular diameters measured with the 37.5~m baseline of the NPOI. The four panels show 
    $\delta$~Cep, its associated check star, $\beta$~Lac, $\eta$~Aql, and its check star 12~Aql.  The 
    diameters are presented as a function of pulsation phase, using the $\delta$~Cep ephemeris of 
    \citet{Mof85} for $\delta$~Cep and $\beta$~Lac, and the $\eta$~Aql ephemeris of \citet{Sza91} for 
    $\eta$~Aql and 12~Aql.  The horizontal dotted lines represent the mean diameters.  The uncertainties for 
    the diameter measurements of the check stars are smaller for two reasons: first, they are larger stars, 
    and hence easier to measure; and second, they are significantly closer to their calibrators than are the 
    Cepheids.}\label{figure:diameters}
\end{figure}

Figure~\ref{figure:diameters} presents the nightly diameters of the Cepheids as functions of pulsation phase 
as calculated from the ephemerides of \citet{Mof85} for $\delta$~Cep and \citet{Sza91} for $\eta$~Aql.  It 
also presents the nightly diameters of the check stars, phased with the same ephemerides as their associated 
Cepheids, to indicate the stability of the diameter measurements for stars of constant diameter.  Note that 
the uncertainties for the check stars are smaller, as expected.


\section{Discussion}\label{sect:discuss}

\subsection{Diameters from the IRFM and the Surface Brightness Technique}\label{sect:diams}

We checked the angular diameters derived here with those derived from the infrared flux method (IRFM) 
\citep{Bla77, Bla90}.  Using the more recent conversions from \citet{Bla94} of $(V-K)_{0}$ to $T_{\rm eff}$ 
and to ``reduced flux'' $\Phi$ (see the equation on their p.~903), we calculated the results summarized in 
Table~3.  The inputs to the IRFM calculation were the intensity mean $V_{0}$ and 
$(V-K)_{0}$ derived from our spline fits to the photometry of \citet{Bar97}, with the dereddening as described 
in \S\ref{sect:reduction}.  For $\delta$~Cep, these values were $3\fm 64$ and $1\fm 36$, respectively, while 
for $\eta~$Aql, they were $3\fm 40$ and $1\fm 49$.


\begin{table}
    \label{table:angdiams}
\begin{tabular}{lll}
    \multicolumn{3}{c}{T{\sc able} 3. A{\sc ngular diameter comparisons (mas)}} \\[10pt]
    \tableline\tableline\\[-8pt]
                        & $\delta$~Cep          & $\eta$~Aql          \\
    \tableline\\[-2pt]
     This work          & 1.520 $\pm$ 0.014     & 1.69 $\pm$ 0.04     \\
     IRFM               & 1.46\phn  $\pm$ 0.07  & 1.77 $\pm$ 0.07     \\
     Surface brightness & 1.50\phn  $\pm$ 0.04  & 1.83 $\pm$ 0.05     \\
     \tableline\\

\end{tabular}
\end{table}

We estimated the overall uncertainty in $\langle \theta_{\rm LD}({\rm IRFM}) \rangle$ for both stars by 
combining the uncertainty of 4\% estimated for this method by \citet{Bla94} with the sensitivity to errors in 
$E_{B-V}$ ($\delta$~Cep: 0.01~mas for a $0\fm 02$ error; $\eta$~Aql: 0.01~mas for a $0\fm 01$ error), to 
errors in $V_{0}$ ($\delta$~Cep: 0.002~mas for a $0\fm 01$ error; $\eta$~Aql: 0.003~mas for a $0\fm 01$ 
error), and to errors in $K_{0}$ (both stars: 0.02~mas for a $0\fm 02$ error).

We also applied the surface-brightness ($S_{\rm V}$) technique, as implemented by \citet{DiB98}, to the 
dereddened \citet{Bar97} photometry (see 
Table~3).  The overall uncertainty in $\langle 
\theta_{\rm LD}(S_{\rm V}) \rangle$ is based on the 2\% uncertainty suggested (but not explicitly stated) by 
\citet{DiB98} and the sensitivity to errors in $E_{B-V}$ ($\delta$~Cep: 0.01~mas for a $0\fm 02$ error; 
$\eta$~Aql: 0.01~mas for a $0\fm 01$ error), in $V_{0}$ (both stars: 0.003~mas for a $0\fm 01$ error), and in 
$K_{0}$ (both stars: 0.02~mas for a $0\fm 02$ error).

The agreement between our $\delta$~Cep results and those of these methods is good: all three diameters agree 
within the uncertainties, although our diameter is larger than the other two.  For $\eta$~Aql, the agreement 
is not quite as good.  Our diameter agrees within the uncertainties with the IRFM diameter, but is slightly 
smaller than the $S_{\rm V}$ diameter, implying a slightly greater surface brightness.

\subsection{Distances from Mean Angular Diameters}\label{sect:mean_diam}

We may use angular diameter measurements to derive the distances to $\delta$~Cep and $\eta$~Aql in two ways.  
The first is to use the mean angular diameter in conjunction with estimates of the linear diameter, which are 
ultimately based on surface brightness techniques.  The second is to use the change in angular diameter in 
conjunction with an estimate of the change in linear diameter derived from radial-velocity measurements.  We 
discuss the first technique here.  As we point out in the next section, the pulsations are only weakly 
detected, so we do not attempt the second method.

If the linear radii of the Cepheids are known from $P$-$R$ relations, the angular diameters can be used to 
calculate distances.  The $P$-$R$ relation of \citet{Lan95} yields $R = 41.5 \pm 1.0 R_{\sun}$ for 
$\delta$~Cep.  Other values for $R$ are similar, e.g., $43.8 \pm 2.9~R_{\sun}$ from the $P$-$R$ relation of 
\citet{Gie99} and $42.4 \pm 4.4~R_{\sun}$ from \citet[CORS]{Cac81}.  \citet{Rip97} present a much different 
result, $52.8~R_{\sun}$ based on a modification of the CORS method.

\citet{Lan95} note that there is a scatter of $\sim 12$\% about the mean radius at fixed period, most of it 
probably due to uncertainties in the radius displacement curve.  If we include that scatter, the 
\citeauthor{Lan95} radius becomes $41.5 \pm 5.1~R_{\sun}$, and the uncertainty is now large enough to include 
the \citet{Gie99} and the \citet{Cac81} results.  Such a mean radius, combined with our mean angular diameter 
of $1.52 \pm 0.014$~mas, yields a distance $D$ of $254 \pm 30$~pc, a distance modulus $(m-M)$ of $7.02 \pm 
0.26$, and an absolute magnitude $\langle M_{\rm V}\rangle = -3.38 \pm 0.26$.  These results are all 
consistent with the Hipparcos parallax, $3.32 \pm 0.58$~mas.

For $\eta$~Aql ($\langle\theta_{\rm LD}\rangle = 1.69 \pm 0.04$~mas), the \citeauthor{Lan95} radius becomes $R 
= 51.6 \pm 5.6 R_{\sun}$.  The \citet{Gie99} relation gives $53.5 \pm 3.7~R_{\sun}$; the \citet{Cac81} 
relation gives $52.0 \pm 5.8~R_{\sun}$; and \citet{Rip97} give $56.7~R_{\sun}$.  Since the enlarged 
\citeauthor{Lan95} uncertainty includes the other values, we use their radius to obtain $D = 284 \pm 31$ and 
$(m-M) = 7.26 \pm 0.23$, which gives $\langle M_{\rm V}\rangle = -3.86 \pm 0.23$, where again we have used the 
reddening described in \S\ref{sect:obs}.  These too are consistent with the Hipparcos parallax, $2.78 \pm 
0.91$~mas.

The absolute magnitudes derived from $R$ and $\langle\theta_{\rm LD}\rangle$ can be compared to those derived 
from the $P$-$L$ relation of \citet{Gie98}, which are $-3\fm 31 \pm 0\fm 04$ for $\delta$~Cep and $-3\fm 66 
\pm 0\fm 04$ for $\eta$~Aql, and the values derived from the $P$-$L$ relation of \citet{Fea98} which are 
$-3\fm 48 \pm 0\fm 11$ and $-3\fm 84 \pm 0\fm 11$.  The value for $\delta$~Cep lies between the two estimates, 
while the result for $\eta$~Aql is closer to that of Feast, Pont, \& Whitelock.  Since the intrinsic scatter is 
probably $\sim 20$\% \citep{Gie93}, and since these two Cepheids are close together in period, our results do 
not distinguish between these two $P$-$L$ relations.

\section{Do We See Pulsations?}\label{sect:pulse}

If we could determine the pulsation amplitudes $\theta_{\rm LD}$ from the present data, we could use the 
displacement derived from radial-velocity curves to make a direct determination of the distances to the 
Cepheids, rather than using $\langle\theta_{\rm LD}\rangle$ and radius estimates.  However, the evidence of 
pulsations in our results is weak, which is not surprising since the fringe spacing of the longest baseline 
used here is 3.9~mas at the mean observing wavelength of $\lambda 700$~nm.

To evaluate the evidence for pulsations, we fit the nightly diameters for each Cepheid with a pulsation shape 
$\delta R(\phi)$ derived from integrating the radial velocity curve using a constant projection factor $p$.  
The appropriate value of $p$ depends in part on the type of radial velocity data used \citep{Hin86}.  For 
$\delta$~Cep, we used $p = 1.31$ with the velocity data of \citet{Sha58}, while for $\eta$~Aql, we used $p = 
1.33$ with the velocity data of \citet{But96}.

With the shape held fixed, there are three quantities to adjust: the mean linear radius, the amplitude of the 
angular diameter variation, and the phase.  We parameterized these three with $\langle\theta_{\rm LD}\rangle$, 
$D$, and $\Delta\phi_v$, the phase shift of $\delta R(\phi)$ such that a positive value of $\Delta\phi_v$ 
corresponds to $R_{\rm min}$ preceding the minimum $V$ magnitude.  We fit the models using five scenarios: (1) 
varying $\theta$, $D$, and $\Delta\phi_v$ simultaneously; (2) holding $\Delta\phi_v$ fixed at zero; (3) 
holding $D$ fixed at 301~pc for $\delta$~Cep and 360~pc for $\eta$~Aql; (4) holding both $\Delta\phi_v$ and 
$D$ fixed; and (5) constant diameter.  (The fixed distances are $1/\pi_{\rm Hip}$.)

The results of these fits are summarized in 
Table~4.  The $\langle\theta_{\rm LD}\rangle$ 
estimates vary slightly between scenarios because the measurements are not evenly distributed in phase.  (To 
illustrate this effect, imagine that we had four high-quality measurements, three at minimum diameter and one 
at maximum.  The mean diameter for a constant-diameter model would be smaller than the mean diameter for a 
pulsating model.)  The $\delta$~Cep models have 15 to 17 degrees of freedom, $\nu$, and the $\eta$~Aql models 
have eight to ten.  The fact that the best $\chi^{2}_{\rm tot}$ values are about equal to $\nu$ reflects the 
fact that we have estimated the uncertainties from the data (\S\ref{subsect:uncert}).


\begin{table}
    \label{table:models}
\begin{tabular}{lcccl}
    \multicolumn{5}{c}{T{\sc able} 4. C{\sc omparison of pulsation solutions}} \\[10pt]
    \tableline\tableline\\[-8pt]
    Model & $\langle \theta_{\rm LD} \rangle$ (mas)  & $D$ (pc)  &  $\Delta\phi_{v}$  &  $\chi^{2}_{\rm tot}$ \\
    \tableline\\[-2pt]
    \multicolumn{5}{c}{$\delta$ Cep} \\[6pt]
    \tableline\\[-2pt]
     Vary $\theta_{\rm LD}$, $D$, $\phi_{v}$ & $1.523 \pm 0.013$ & 440$^{+330\phn}_{-130}$ & $0.10 \pm 0.06$ & 13.6 \\
     Fix $\phi_{v} = 0$                      & $1.520 \pm 0.014$ & 650$^{+1200}_{-260}$ & ---            & 16.0 \\
     Fix $D = 301$ pc                        & $1.523 \pm 0.014$ & ---                 & $0.10 \pm 0.05$ & 14.6 \\
     Fix $\phi_{v}$ and $D$                  & $1.520 \pm 0.015$ & ---                 & ---             & 19.3 \\
     No pulsation                            & $1.520 \pm 0.015$ & ---                 & ---             & 18.5 \\
     \tableline\\[-2pt]
     \multicolumn{5}{c}{$\eta$ Aql} \\[6pt]
     \tableline\\[-2pt]
     Vary $\theta_{\rm LD}$, $D$, $\phi_{v}$ & $1.688 \pm 0.035$ & 240$^{+190\phn}_{-75}$  & $0.10 \pm 0.06$ & $\phantom{1}$8.0 \\
     Fix $\phi_{v} = 0$                      & $1.695 \pm 0.037$ & 380$^{+600\phn}_{-150}$ & ---             & 10.3 \\
     Fix $D = 360$ pc                        & $1.681 \pm 0.033$ & ---                 & $0.10 \pm 0.06$ & $\phantom{1}$8.5 \\
     Fix $\phi_{v}$ and $D$                  & $1.696 \pm 0.036$ & ---                 & ---             & 10.3 \\
     No pulsation                            & $1.665 \pm 0.039$ & ---                 & ---             & 12.6 \\
    \tableline\\
\end{tabular}
\end{table}


The significance of the differences in $\chi^{2}_{\rm tot}$ can be assessed using an $F$ test, from which we 
infer that the pulsations appear only at the $\sim 1.5\sigma$ to $2\sigma$ level for both $\delta$~Cep and 
$\eta$~Aql.  The sizes of the distance uncertainties also reflect the low significance of the pulsations.  We 
feel that this level of significance is insufficient to claim a clear detection, given the importance of such 
a detection.  The $\chi^{2}_{\rm tot}$ values also imply that any phase shift is also present at only the 
$\sim 1.5 \sigma$ level.  The $\langle \theta_{\rm LD}\rangle$ results that we selected for reporting here are 
those for $\Delta\phi_v = 0$ because the significance of a phase shift in our data is small.

To improve our chances of detecting the pulsations, we need either longer baselines or an improved 
understanding of the calibration of the fringe visibilites.  Longer baselines at the NPOI will become 
available in the near future, while efforts to improve the calibration are currently under way.

\section{Summary}\label{sect:summary}

We have used the NPOI to measure the angular diameters of $\delta$~Cep and $\eta$~Aql over the $\lambda\lambda 
850-600$~nm wavelength range.  The mean angular diameters are $1.520 \pm 0.014$~mas for $\delta$~Cep and $1.69 
\pm 0.04$~mas for $\eta$~Aql.  With the current level of precision in $P$-$R$ and $P$-$L$ relations, the small 
difference in period between these two Cepheids, and the uncertainties in the distances, it is difficult to 
distinguish between different relations.  The distances obtained from the $P$-$R$ radii and our angular 
diameters are slightly smaller than, but still consistent with, the Hipparcos distances.

The evidence for Cepheid pulsations in our results is weak, as one would expect for measurements of diameters 
near the resolution limit of the current configuration of the NPOI. We evaluate the strength of this evidence 
from the $\chi^{2}_{\rm tot}$ of fits to a constant diameter and to pulsating models with pulsation shapes 
derived from integrating radial-velocity curves.  The pulsations are only weakly seen in our data, at the 
$\sim 1.5$ to $2\sigma$ level.

It would be advantageous to measure the angular diameter changes of the Cepheids directly, so that one could 
avoid the problem of color vs.\ surface brightness calibration in determining their distances.  In addition, a 
measurement of the limb darkening---and changes therein---would add confidence to the knowledge of the 
projection factor $p$ used to convert radial velocity into pulsation velocity in any such analysis.  We expect 
that a 64~m baseline for the NPOI will be available by the end of 2000.  The increased resolution will make 
possible the unambiguous direct detections of the pulsations of $\delta$~Cep, $\eta$~Aql, and $\zeta$~Gem, 
three of the four brightest Cepheids in the sky.

\acknowledgements 
Several improvements to our presentation are due to the comments of the anonymous referee.  The NPOI is funded 
by the Office of Naval Research and the Oceanographer of the Navy.  This research made use of the SIMBAD 
database operated by the CDS, Strasbourg, France, the {\it Database of Galactic Classical Cepheids} 
(\url{http://ddo.astro.utoronto.ca/cepheids.html}) maintained at the David Dunlap Observatory of the 
University of Toronto \citep{Fer95}, and the {\it Cepheid Photometry and Radial Velocity Data Archive} 
(\url{http://physun.physics.mcmaster.ca/Cepheid/}) maintained at McMaster University \citep{Wel99}.





\newpage
\begin{thebibliography}{}
    
    \bibitem[Armstrong et al.(1998)]{Arm98}Armstrong, J.T., et al.  1998, \apj, 496, 550 
    \bibitem[Barnes et al.(1997)]{Bar97}Barnes, T.G. III, Fernley, J.A., Frueh, M.L., Garc\'ia Navas, J., Moffett, J.J., \& 
    Skillen, I.  1997, \pasp, 109, 645
    \bibitem[Blackwell \& Lynas-Gray(1994)]{Bla94}Blackwell, D.E., \& Lynas-Gray, A.E.  1994, \aap, 282, 899
    \bibitem[Blackwell et al.(1990)]{Bla90}Blackwell, D.E., Petford, A.D., Arribas, S., Haddock, K.J., \& Selby, 
    M.J.  1990, \aap, 232, 396
    \bibitem[Blackwell \& Shallis(1977)]{Bla77}Blackwell, D.E., \& Shallis, M.J.  1977, \mnras, 180, 177
    \bibitem[Brown et al.(1974)]{Bro74} Brown, R.\ Hanbury, Davis, J., Lake, R.J.W., \& Thompson, R.J.
    1974, \mnras, 167, 475 
    \bibitem[Butler(1993)]{But93}Butler, R.P.  1993, \apj, 415, 323
    \bibitem[Butler, Bell, \& Hindsley(1996)]{But96}Butler, R.P., Bell, R.A., \& Hindsley, R.B.  1996, \apj, 
    461, 362
    \bibitem[Caccin et al.(1981)]{Cac81} Caccin, B., Onnembo, A., Russo, G., \& Sollanzo, C.\ (CORS) 1981, \aap, 97, 104
    \bibitem[Di Benedetto(1998)]{DiB98}Di Benedetto, G.P.  1998, \aap, 339, 858
    \bibitem[ESA(1997)]{ESA97}ESA 1997, The Hipparcos Catalogue, ESA SP-1200
    \bibitem[Evans(1976)]{Eva76}Evans, N.R. 1976, \apjs, 32, 399
    \bibitem[Evans \& Jiang(1993)]{Eva93}Evans, N.R., \& Jiang, J.H.  1993, \aj, 106, 726
    \bibitem[Feast, Pont, \& Whitelock(1998)]{Fea98}Feast, M.W., Pont, F., \& Whitelock, P. 1998, \mnras, 298, L43
    \bibitem[Fernie et al.(1995)]{Fer95}Fernie, J.D., Beattie, B., Evans, N.R., and Seager, S. 1995, IBVS 
    No.~4148; http://ddo.astro.utoronto.ca/cepheids.html
    \bibitem[Gieren, Fouqu\'e, \& G\'omez(1998)]{Gie98}Gieren, W.P., Fouqu\'e, P., \& G\'omez, M. 1998, \apj, 496, 17
    \bibitem[Gieren, Barnes, \& Moffett(1993)]{Gie93}Gieren, W.P., Barnes, T.G.\ III, \& Moffett, T.J.  
    1993, \apj, 418, 135
    \bibitem[Gieren, Moffett, \& Barnes(1999)]{Gie99}Gieren, W.P., Moffett, T.J., \& Barnes, T.G.\ III, 1999, \apj, 512, 533
    \bibitem[Hindsley \& Bell(1986)]{Hin86}Hindsley, R.B., \& Bell, R.A.  1986, \pasp, 98, 881
    \bibitem[Hindsley \& Bell(1989)]{Hin89}Hindsley, R.B., \& Bell, R.A.  1989, \apj, 341, 1004
    \bibitem[Jacobsen \& Wallerstein(1981)]{Jac81}Jacobsen, T.T., \& Wallerstein, G. 1981, \pasp, 93, 481
    \bibitem[Laney \& Stobie(1995)]{Lan95}Laney, C.D., \& Stobie, R. S. 1995, \mnras, 274, 337
    \bibitem[McWilliam(1990)]{McW90}McWilliam, A.  1990, \apjs, 74, 1075
    \bibitem[Moffett \& Barnes(1985)]{Mof85}Moffett, T.J., \& Barnes, T.G.\ III  1985, \apjs, 58, 843
    \bibitem[Mourard et al.(1997)]{Mou97}Mourard, D., et al. 1997, \aap, 317, 789 
    \bibitem[Nordgren et al.(1999)]{Nor99}Nordgren, T.E., et al.  1999, \aj, 118, 3032
    \bibitem[Nordgren et al.(2000)]{Nor00}Nordgren, T.E., Armstrong, J.T., Germain, M.E., Hindsley, R.B., 
    Hajian, Arsen R., Sudol, J.J., \& Hummel, C.A.  2000, \apj, in press
    \bibitem[Ripepi et al.(1997)]{Rip97}Ripepi, V., Barone, F., Milano, L., \& Russo, G. 1997, \aap, 318, 797
    \bibitem[Shane(1958)]{Sha58}Shane, W.W.  1958, \apj, 127, 573
    \bibitem[Szabados(1991)]{Sza91}Szabados, L.  1991, Comm.\ Konkoly Obs., 96, 138
    \bibitem[Van Hamme(1993)]{VHa93}Van Hamme, W.  1993, \aj, 106, 2096 
    \bibitem[Welch(1999)]{Wel99}Welch, D.  1999, Cepheid Photometry \& Radial Velocity Data Archive, 
    http://physun.physics.mcmaster.ca/Cepheid 

\end{thebibliography}
\end{document}